\documentclass{article}
\usepackage{natbib}
\usepackage{graphicx}

\begin{document}
Research Note, submitted to the Canadian Journal of Remote Sensing

\subsection*{Detection of coherent reflections with GPS bipath interferometry}

Achim Helm, Georg Beyerle and Markus Nitschke\\
GeoForschungsZentrum Potsdam, Dept. Geodesy \& Remote Sensing, Potsdam, Germany\\
{\footnotesize
Corresponding author address:
Achim Helm, GeoForschungsZentrum Potsdam,
Dept. Geodesy \& Remote Sensing, Telegrafenberg, D-14473 Potsdam, Germany.
Tel.: +49-331-288-1812; fax: +49-331-288-1111. E-mail: helm@gfz-potsdam.de
}

\bigskip

{\bf Abstract }{Results from a GPS reflectometry experiment  with a 12 channel ground-based GPS receiver above
two lakes in the Bavarian Alps are presented.
The receiver measures in open-loop mode the coarse/aquisition code correlation function
of the direct and the reflected signal of one GPS satellite simultaneously.
The interference between the coherently reflected signal and a model signal,
which is phase-locked to the direct signal, causes variations in the amplitude of the
in-phase and quad-phase components of the correlation sums.
From these amplitude variations the relative altimetric height is determined within a precision of 2~cm.}




\section{Introduction}
Satellite-based active altimeters on ENVISAT and JASON deliver
valuable ocean height data sets for global climate modelling. In order to
improve the climate models, altimetric data of higher resolution in space and
time is required. This gap can potentially be filled with GPS-based altimetric
measurements.
Additionally, ground-based GPS receivers can monitor ocean heights
in coastal areas where satellite altimetry data get coarse and decrease in quality 
\citep{fu2001,shum1997}.

Since GPS altimetry has been proposed as a novel remote sensing capability
\citep{martin-neira1993}, many studies have been carried out at
different observation heights and platforms.
While Earth-reflected GPS signals
have been observed from spaceborne instruments \citep{lowe2002a,beyerle2002}
and the CHAMP and SAC-C satellites already are equipped with
Earth/nadir looking GPS antennas, work is in progress in order to
establish satellite-based GPS altimetry \citep{hajj2003}.
Airborne campaigns have been conducted (e.g. \citet{garrison1998},
\citet{garrison2000}, \citet{rius2002})
 and recently reached a 5-cm height precision \citep{lowe2002c}.
Ground-based GPS altimetry measurements have been performed at
laboratory scale of some meters height with 1-cm height precision \citep{martin-neira2002} up to
low-altitudes height (e.g. \citet{anderson2000}, \citet{martin-neira2001}) and
reached a 2-cm height precision \citep{treuhaft2001}.

In this study a 12 channel GPS receiver is used \citep{kelley2002}, that
has been extended with a coarse/acquisition (C/A) code correlation function tracking mode.
In this coherent delay mapping (CDM) mode the direct GPS signal is tracked
while concurrently the reflected signal is registered in open-loop mode.
Using the L1 carrier phase the relative altimetric height is determined from the
components of the reflected signal.

\section{Experimental Setup and Data Acquisition}
The experiment was conducted on 8 - 10 July 2003, 50 km south of Munich, Germany, 
in the Bavarian alpine upland at
the mountain top of Fahrenberg (47.61$^\circ$N, 11.32$^\circ$E) at a height of about 1625 m asl.
Mount Fahrenberg belongs to the Karvendel mountains and from the mountain top
unobstructed view is available to lake Kochelsee (surface area about 6 km$^2$) to the north
and lake Walchensee (surface area about 16 km$^2$) to the south.
Following a schedule of predicted GPS reflection events, the receiver antenna was turned towards
the lake surface of Kochelsee (about 599 m asl) or Walchensee (about 801 m asl).
The antenna was tilted about $45^\circ$ towards the horizon.
During a GPS reflection event the direct and the reflected signals interfere at the antenna center
(e.g. \citet{parkinson1996}).
The interference causes amplitude fluctuations that are quantitatively analyzed
to determine the height variation of the specular reflection point.

\begin{figure}
\includegraphics[width=0.7\textwidth]{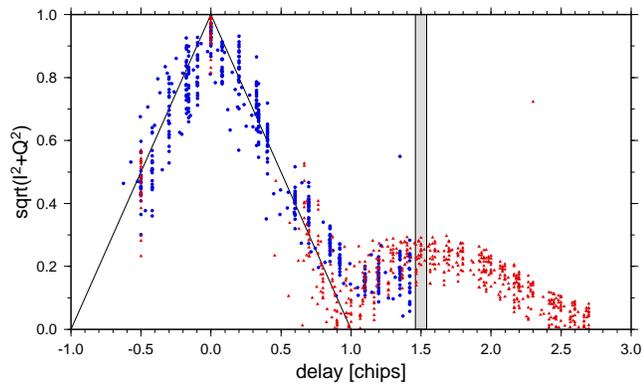}
 \caption{Delay mapped waveform of a reflection event (PRN 16) at 1334:17 UTC 8 July 2003,
 antenna oriented towards Kochelsee.
 The delay is given in relation to the maximum peak of the direct signal.
  Blue circles and red triangles indicate
 2 measurements (0.5-second duration) starting 120 (blue) and 267 (red) seconds after the
 start of the measurement.
 In the second case (red) the 2-chip-wide interval of covered chip code offsets is centered
 at the maximum of the reflected signal.
 The points reveal the measured waveform of the direct and reflected correlation signal.
 The thin black triangle marks the theoretical C/A code correlation function of the direct signal.
 The grey box marks
 the maximum of the reflected signal.}\label{f-waveform}
\end{figure}

The receiver is based on the OpenSource~GPS design
\citep{kelley2002} and was modified to allow for
open-loop tracking of reflected signals.
The receiving antenna is an active RHCP patch antenna
(AT575-70C from AeroAntenna Technology Inc.)
with 4 dBic gain, 54~mm in diameter and a hemispheric field-of-view.
Operating in CDM mode all 12~correlator channels are tuned to the
same GPS satellite by setting the appropriate pseudo-random noise (PRN) value.
The correlation between the received and model (replica) signal is realized in hardware
(Zarlink GP2021, \citet{zarlink2001}).
While one channel (the master channel) continues to track
the direct signal,
the carrier and code tracking loops of the 11~remaining
channels (slave channels) are synchronized to the master
channel.
Each channel consists of the prompt and the early tracking arm
at zero and at -0.5~chip code delay, respectively.
Thus, $2\times11=22$ delays are available to map
the C/A code correlation function of the reflection signature.
In CDM mode the slave carrier and code phase-locked
loops (PLLs) are opened and their loop feed-back is obtained
from the master PLL.
All carrier loops operate with zero carrier phase offset
with respect to the master channel;
in the code loop feed-back, however, delays covering an interval of
2~chips (about 2~$\mu$s) with a step size of 0.1~chips
(about 100~ns) are inserted.
In-phase and quad-phase correlation-sums of each channel are summed
over 20~ms taking into account the navigation bit
boundaries and stored together with code and carrier
phases to hard disk at a rate of 50 Hz.
Figure \ref{f-waveform} illustrates the CDM mode:
while the direct GPS signal is tracked with the prompt and early arm
of the master channel at 0 and -0.5 chips code offset, the prompt and
early arms of the remaining 11 slave channels are set to chip code offsets
between 0.4 and 2.7 to map the reflected signal
(corresponding to an optical
path difference of 120 to 810~m).
In Figure \ref{f-waveform} the root-sum-squared in-phase and quad-phase values of the reflected signal are
plotted as a function of code delay. The maximum power of the reflected signal is about
$20\log 0.2=-14$ dB below the direct signal's power.
The peak of the correlation function is separated by a delay of 1.5 chips from the
direct signal's correlation peak.

\begin{figure}
\includegraphics[width=0.7\textwidth]{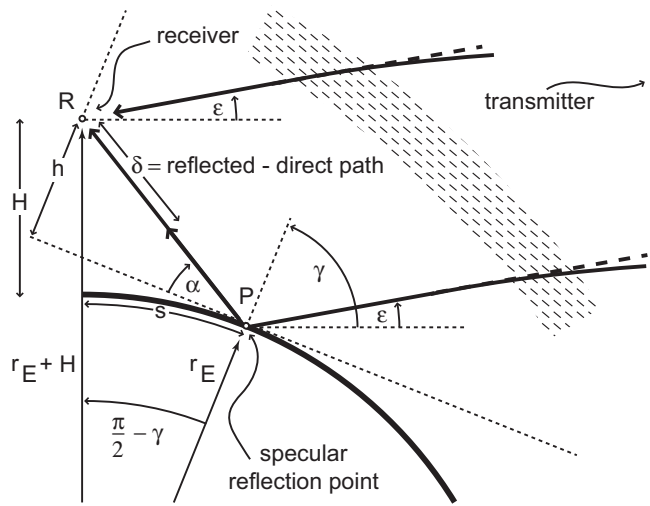}
 \caption{Geometry used to express the observed path difference $\delta$ in terms of the known
 receiver position R with height H and the GPS satellite elevation angle $\epsilon$ and the calculated position of
  the specular reflection point P. $h$ denotes the height variations normal to the tangential plane at P.
  Note that $\epsilon$ has to be corrected by $\Delta\epsilon_{tropo}$ due to the bending effect
  caused by the Earth's troposphere.}\label{f-geometry}
\end{figure}

\begin{figure}
\includegraphics[width=0.7\textwidth]{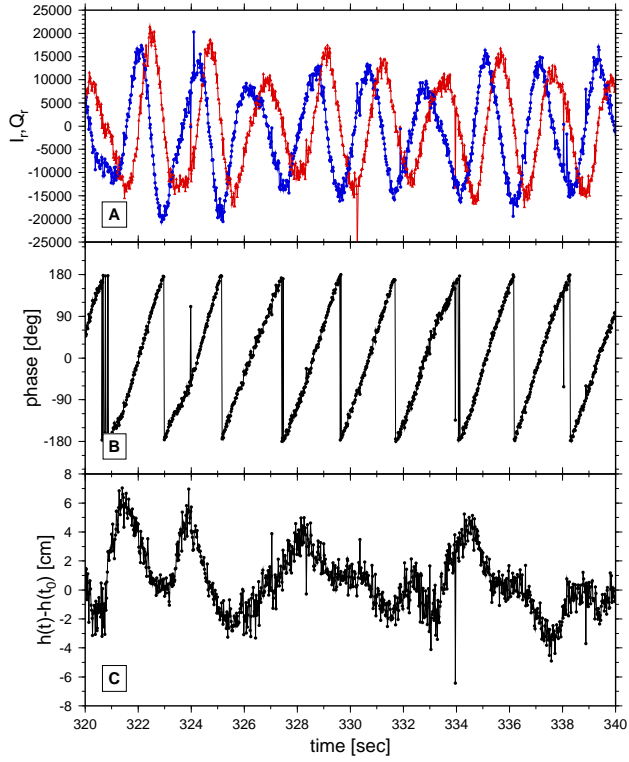}
 \caption{Panel A shows the demodulated reflected in- and quad-phase data
 $\tilde{I}_r$ (blue circles) and $\tilde{Q}_r$ (red triangles) (PRN 16, elevation from $11.04^\circ$ to $10.99^\circ$),
 antenna oriented towards Kochelsee at 1334:17 UTC 8 July 2003, as a function of time since measurement start.
 With Eq. \ref{e-phi} the phase $\phi$  (Panel B) and
 from Eq. \ref{e-ht} and \ref{e-h0} the relative height $h(t)-h(t_0)$ is
calculated (Panel C), with $H(t_0)=1022.5$ m.}\label{f-iqdata}
\end{figure}

Data analysis is performed in the following way:
first, the code delay corresponding to the maximum of the reflected waveform is determined.
Second, all in-phase and quad-phase correlation sum  values $I_r$ and $Q_r$ are extracted from the
raw data which lie within
a certain delay interval (grey box in Figure \ref{f-waveform}) around the maximum code delay.
The navigation message is demodulated according to
\begin{eqnarray}
\tilde{I}_r & = & \mbox{sign}(I_d)\,I_r \nonumber \\
\tilde{Q}_r & = & \mbox{sign}(I_d)\,Q_r,
 \end{eqnarray}
where $I_d$ denotes the in-phase value of the master channel.
Figure \ref{f-iqdata} A shows the oscillations of $\tilde{I}_r$ and $\tilde{Q}_r$
caused by the interference between the reflected and the replica GPS signal.
The phasor $\tilde{I}_r + i\,\tilde{Q}_r$ rotates by about $+\,$0.5 Hz due to the
decreasing path length difference between the direct and the reflected signal,
since during this measurement the GPS satellite moved towards the horizont.
Note the phase offset of $90^\circ$ between $\tilde{I}_r$ and $\tilde{Q}_r$.
The phase $\phi$ (Fig. \ref{f-iqdata} B) is calculated from the four quadrant arctangent
\begin{equation}
	\phi   = \mbox{atan2}(\tilde{Q}_r,\tilde{I}_r)\label{e-phi}
 \end{equation}
and is unwrapped by adding $\pm\,2\pi$ when the difference between
consecutive values exceeds $\pi$, resulting in the accumulated phase $\phi_{a}$.
The optical path length difference $\delta$ between direct and reflected signal is calculated
from the accumulated phase $\phi_{a}$ and the L1 carrier wavelength $\lambda_{L1}=0.1903$ m at the observation time $t$
\begin{equation}
  \delta(t) = \frac{\phi_{a}(t)}{2\pi}\, \lambda_{L1}.
\end{equation}
Starting with a height estimate $H(t_0)$, the temporal evolution of the altimetric height
variation $h(t)-h(t_0)$, normal to the tangent plane at the reflection point P, is
calculated from \citep{treuhaft2001}
\begin{eqnarray}
h(t)&=& \frac{\delta(t)-\delta(t_0)+2\,h(t_0)\,\sin\alpha(t_0)}{2\,\sin\alpha(t)}\label{e-ht}\\
h(t_0) &=& (H(t_0)+r_E)\cos\frac{s}{r_E}-r_E \label{e-h0}
\end{eqnarray}
with the arclength $s$ defined in Figure \ref{f-geometry}, an Earth radius $r_E=6371$ km and
\begin{eqnarray}
\alpha&=& \epsilon + \frac{\pi}{2} - \gamma\\\label{e-alpha}
\epsilon&=& \epsilon_{eph} + \Delta\epsilon_{tropo},\label{e-epsilon}
 \end{eqnarray}
assuming an infinite distance to the GPS transmitter.
$\epsilon_{eph}$ is calculated
from the broadcast ephemeris data \citep{gps_sps1995}, the correction $\Delta\epsilon_{tropo}$
accounts for refraction caused by atmospheric refractivity.
The tropospheric correction is derived from a geometric raytracing calculation
using a refractivity profile obtained from meteorological analyse provided by the
European Centre for Medium-Range Weather Forecasts.
The position of the specular reflection point P as function of $\gamma$ (Figure \ref{f-geometry})
is calculated following \citet{martin-neira1993}.

Thus, the altimetric height change of the GPS receiver above the reflecting
surface is determined from the carrier phase difference between the direct and reflected 
GPS signal (Figure \ref{f-iqdata} C).

\section{Data Analysis and Discussion}
During all 3 days several reflection events were observed from both lake surfaces with different GPS
satellites at elevation angles between about $10^\circ$ - $15^\circ$,
indicated by a clearly visible waveform (see Figure \ref{f-waveform}).
Several outliers can be observed in the data records.
Most likely they are caused by overheating of the hardware correlator chip
[S. Esterhuizen, University of Colorado, personal communication, 2003].
In this study outliers are removed in the following way: a value is calculated by 
linear extrapolation from the last 3 values of $\tilde{I}_r(t)$.
If the difference between extrapolated and actual value exceeds a threshold (20000-22000),
the extrapolated value is taken.
The same is applied to the $\tilde{Q}_r(t)$ data.
Additionally cycle slips (sporadic height jumps of about $\lambda_{L1}$~m in adjacent data points)
can be observed in the optical path length difference $\delta(t)$.
The distortion of the data by cycle slips could be minimized by applying the same method as
above to $\delta(t)$.
Continuous data segments without cycle slips are chosen for height change determination.
The mean receiver height above the lake surface is
not expected to change during the short analyzed time periods.
From topographic maps (scale 1:25000, Bayerisches Landesvermessungsamt, 1987) the heights $H(t_0)$ are estimated to be 1026 m $\pm\,$5 m (Kochelsee)
and 824 m $\pm\,$5 m (Walchensee), respectively.
By minimization of the linear trend of $h(t)-h(t_0)$ we obtain a $H(t_0)$ of 1022.5 m (Kochelsee)
and 827.5 m (Walchensee).

\begin{figure*}
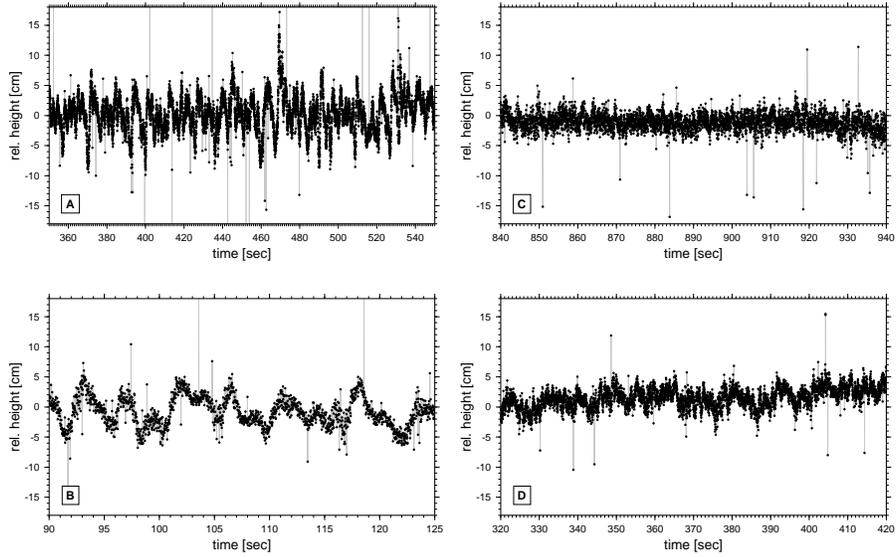

\includegraphics[width=0.495\textwidth]{helm_fig4a.eps}\includegraphics[width=0.495\textwidth]{helm_fig4c.eps}
\includegraphics[width=0.495\textwidth]{helm_fig4b.eps}\includegraphics[width=0.495\textwidth]{helm_fig4d.eps}
\caption{The left panels show relative height measurements at Kochelsee (PRN 16),
starting at 1334:17 UTC 8 July 2003 (Panel A) and starting at 1327:17 UTC 10 July 2003 (Panel B)
as a function of time since the start of the observation. PRN 16 changed elevation from $11.0^\circ$ to
$10.4^\circ$ (Panel A) and from $11.4^\circ$ to $11.3^\circ$ (Panel B).
On the right panels height measurements at Walchensee are shown,
PRN 20 (elevation from $14.7^\circ$ to $14.1^\circ$), starting at 1257:15 UTC 8 July 2003 (Panel C)
and PRN 11 (elevation from $14.3^\circ$ to $13.6^\circ$), starting at 1110:21 UT 9 July 2003 (Panel D).
$H(t_0)=1022.5$ m (Kochelsee) and $H(t_0)=827.5$ m (Walchensee).
}\label{f-rhdata}
\end{figure*}

Figure \ref{f-rhdata} A and B plot the relative height change between the receiver and the
reflection point at the surface of lake Kochelsee.
Both observations used the same PRN, but were taken on different days.
The height varies within an interval of about $\pm\,5$~cm with a
standard deviation of about 3.1 and 2.6~cm.
Figure \ref{f-rhdata} C and D show the height changes between the receiver and the
reflection point at the surface of lake Walchensee.
Again both observations were taken on different days and used different PRNs.
Compared to the Kochelsee data, the height varies within a height interval of about $\pm\,2.5$~cm with a
standard deviation of about 1.4 and 1.7~cm.

The different height variations at both lakes can be explained by different local
wind and wave height conditions.
As lake Walchensee is completely surrounded by mountains, waves are mainly driven by
local, thermal induced winds which mainly occur at noon.
Lake Kochelsee is open to the north, so longer lasting wind can build up waves
on the lake surface.

\section{Conclusions and Outlook}
Open-loop tracking of the reflected signals allows the determination of
the relative altimetric height with 2-cm precision.
Different height changes can be observed at Kochelsee and Walchensee which reflect the
different wind and wave height conditions at the two lakes.
The relationship between the observed height changes and wind speed
(e.g. \citet{caparrini1998}, \citet{lin1999}, \citet{komjathy2000a}, \citet{zuffada2003a}, 
\citet{cardellach2003b}) will be subject of further studies.
The present receiver implementation is limited to the observation of one GPS satellite at a time.
To fully use the potential of GPS reflections the receiver will be modified to
keep track of several GPS reflections simultaneously.

Our results suggest that open-loop tracking is possible with low-gain and wide field-of-view antennas, showing the
potential of this method also for space-based measurements of GPS reflections.

\begin{footnotesize}

\section*{Acknowledgments}
This work would not have been possible without the open source projects
OpenSource~GPS and RTAI-Linux.
We thank Clifford Kelley  and the developers of RTAI-Linux for making their work available.
Helpful discussion with Robert Treuhaft and Philipp Hartl are gratefully acknowledged.
We thank T. Schmidt, C. Selke and A. Lachmann for their help and technical support.
The ECMWF provided meteorological analysis fields.



\section*{List of Symbols}

\begin{itemize}
\item[$R$] receiver position
\item[$P$] specular reflection point position
\item[$\epsilon$] elevation angle of the GPS satellite above local horizon plane at R
\item[$\delta$] observed path difference between direct and reflected signal path
\item[$r_E$] Earth radius
\item[$H$] receiver height
\item[$h$] height variations normal to tangential plane at P
\item[$\alpha$] angle of reflection above tangential plane at P
\item[$\gamma$] angle between normal of tangential plane and local horizon plane at P
\item[$s$] arc length from subreceiver point to specular reflection point P
\item[$I_d$] in-phase correlation sum of the direct data
\item[$I_r$] in-phase correlation sum of the reflected data
\item[$Q_r$] quad-phase correlation sum of the reflected data
\item[$\tilde{I}_r$] $I_r$ demodulated from navigation message
\item[$\tilde{Q}_r$] $Q_r$ demodulated from navigation message
\item[$\phi$] phase
\item[$\phi_a$] accumulated phase
\item[$\lambda_{L1}$] L1 carrier wavelength
\item[$\epsilon_{eph}$] elevation angle calculated from broadcast ephemeris data
\item[$\Delta\epsilon_{tropo}$] tropospheric correction to elevation angle
\end{itemize}
\end{footnotesize}

\end{document}